\documentclass[prd,preprint,superscriptaddress,showpacs,byrevtex]{revtex4}
\usepackage{bm}
\def\fsl#1{\setbox0=\hbox{$#1$}           
   \dimen0=\wd0                                 
   \setbox1=\hbox{/} \dimen1=\wd1               
   \ifdim\dimen0>\dimen1                        
      \rlap{\hbox to \dimen0{\hfil/\hfil}}      
      #1                                        
   \else                                        
      \rlap{\hbox to \dimen1{\hfil$#1$\hfil}}   
      /                                         
   \fi}                                         %
\newcommand{\be}{\begin{equation}}
\newcommand{\ee}{\end{equation}}
\newcommand{\bea}{\begin{eqnarray}}
\newcommand{\eea}{\end{eqnarray}}
\newcommand{\beq}{\begin{equation}}
\newcommand{\eeq}{\end{equation}}
\newcommand{\beqs}{\begin{eqnarray}}
\newcommand{\eeqs}{\end{eqnarray}}

\begin{document}
\title{ Fragmentation Function in Non-Equilibrium QCD Using Closed-Time Path Integral Formalism }
\author{Gouranga C Nayak } \email{gnayak@uic.edu}
\affiliation{Department of Physics, University of Illinois, Chicago, IL 60607, USA }
\begin{abstract}
In this paper we implement Schwinger-Keldysh closed-time path
integral formalism in non-equilibrium QCD to the definition of Collins-Soper fragmentation function.
We consider a high $p_T$ parton in QCD medium at initial time $\tau_0$ with arbitrary
non-equilibrium (non-isotropic) distribution function $f(\vec{p})$ fragmenting to hadron.
We formulate parton to hadron fragmentation function in non-equilibrium QCD in the
light-cone quantization formalism. It may be possible to include final state interactions
with the medium via modification of the Wilson lines in this definition of the non-equilibrium
fragmentation function. This may be relevant to study hadron production
from quark-gluon plasma at RHIC and LHC.
\end{abstract}
\pacs{ PACS: 12.39.St, 13.87.Fh, 13.87.Ce, 13.85.Ni }
\maketitle
%
%
\pagestyle{plain}
\pagenumbering{arabic}
\section{Introduction}
RHIC and LHC heavy-ion colliders are the best facilities to study quark-gluon
plasma in the laboratory. Since two nuclei travel almost at speed of light,
the QCD matter formed at RHIC and LHC may be in non-equilibrium. In order
to make meaningful comparison of the theory with the experimental data on hadron
production, it may be necessary to study nonequilibrium-nonperturbative QCD at
RHIC and LHC. This, however, is a difficult problem.

Non-equilibrium quantum field theory can be studied by using Schwinger-Keldysh closed-time
path (CTP) formalism \cite{schw,keldysh,cooper1}.
However, implementing CTP in non-equilibrium at RHIC and LHC is a very difficult problem,
especially due to the presence of gluons in non-equilibrium and hadronization etc.
Recently, one-loop resumed gluon propagator in non-equilibrium
in covariant gauge is derived in \cite{greiner,cooper}.

High $p_T$ hadron production at high energy $e^+e^-$, $ep$ and $pp$ colliders is studied
by using Collins-Soper fragmentation function \cite{collins,frag}.
For a high $p_T$ parton fragmenting to hadron, Collins-Soper derived an expression
for the fragmentation function based on field theory and factorization properties in QCD at
high energy \cite{george}. This fragmentation function is universal in the sense that, once
its value is determined from one experiment it explains the data at other experiments.

The derivation of parton to hadron fragmentation
function in QCD medium based on first principle calculation is not done so far. This
can be relevant at RHIC and LHC heavy-ion colliders to study hadron production
from quark-gluon plasma. Further complication arises because the partons at
RHIC and LHC may be in non-equilibrium.

In this paper we note that, one can implement closed-time path integral formalism
in non-equilibrium QCD to the definition of Collins-Soper
fragmentation function. We consider a high $p_T$ parton in medium at
initial time $\tau_0$ with arbitrary non-equilibrium (non-isotropic) distribution function $f(\vec{p})$
fragmenting to hadron. We formulate parton to hadron
fragmentation function in non-equilibrium QCD in the light-cone quantization formalism.
The special case $f(\vec{p})=\frac{1}{e^{\frac{p_0}{T}}\pm 1}$ corresponds to the finite
temperature QCD in equilibrium. This fragmentation function may be relevant to study hadron production from quark-gluon plasma at RHIC and LHC.

We find the following definition of the parton to hadron fragmentation function
in non-equilibrium QCD by using closed-time path integral formalism.
For a quark ($q$) with arbitrary non-equilibrium distribution function $f_q(\vec{k})$ at
initial time, the quark to hadron fragmentation function is given by
\bea
&& D_{H/q}(z,P_T)
= \frac{1}{2z~[1+f_q({\vec k})]} \int dx^- \frac{d^{d-2}x_T}{(2\pi)^{d-1}}  e^{i{k}^+ x^- + i {P}_T \cdot x_T/z} \nonumber \\
 &&~\frac{1}{2}{\rm tr_{Dirac}}~\frac{1}{3}{\rm tr_{color}}[\gamma^+<in| \psi(x^-,x_T) ~\Phi[x^-,x_T]~a^\dagger_H(P^+,0_T)  a_H(P^+,0_T) ~\Phi[0]~{\bar \psi}(0)  |in>]~~~~~
\label{qnf}
\eea
where $z$ (=$\frac{P^+}{k^+}$) is the longitudinal momentum fraction of the hadron with respect to the
parton and $P_T$ is the transverse momentum of the hadron. $|in>$ is the initial state of the
non-equilibrium QCD medium in the Schwinger-Keldysh $in-in$ closed-time path formalism.

For a gluon ($g$) with arbitrary non-equilibrium distribution function $f_g(\vec{k})$ at
initial time, the gluon to hadron fragmentation function is given by
\bea
&& D_{H/g}(z,P_T)
= \frac{1}{2zk^+~[1+f_g({\vec k})]} \int dx^- \frac{d^{d-2}x_T}{(2\pi)^{d-1}}  e^{i{k}^+ x^- + i {P}_T \cdot x_T/z} \nonumber \\
&&~\frac{1}{8} \sum_{a=1}^8 [<in| F^{+\mu}_a(x^-,x_T) ~\Phi[x^-,x_T]~a^\dagger_H(P^+,0_T)  a_H(P^+,0_T)~\Phi[0]~  F^+_{\mu a}(0) |in>].
\label{gnf}
\eea

The path ordered exponential
\bea
\Phi[x^\mu ]_{ab}={\cal P}~ {\rm exp}[ig\int_{0}^\infty d\lambda~ n \cdot A^c(x^\mu +n^\mu \lambda )~T^c_{ab}]
\label{wilf}
\eea
is the Wilson line \cite{tucci}. It may be possible to include final state interactions
with the medium via modification of the Wilson lines in this definition of the non-equilibrium
fragmentation function, similar to the $p_T$ distribution of the
parton distribution function studied in \cite{beli}.

Eqs. (\ref{qnf}) and (\ref{gnf}) can be compared with the following definition of Collins-Soper
fragmentation function \cite{collins}:
\bea
&& D_{H/q}(z,P_T)
= \frac{1}{2z} \int dx^- \frac{d^{d-2}x_T}{(2\pi)^{d-1}}  e^{i{k}^+ x^- + i {P}_T \cdot x_T/z} \nonumber \\
 &&~\frac{1}{2}{\rm tr_{Dirac}}~\frac{1}{3}{\rm tr_{color}}[\gamma^+<0| \psi(x^-,x_T) ~\Phi[x^-,x_T]~a^\dagger_H(P^+,0_T)  a_H(P^+,0_T) ~\Phi[0]~{\bar \psi}(0)  |0>]
\label{qvf}
\eea
and
\bea
&& D_{H/g}(z,P_T)
= -\frac{1}{2zk^+} \int dx^- \frac{d^{d-2}x_T}{(2\pi)^{d-1}}  e^{i{k}^+ x^- + i {P}_T \cdot x_T/z} \nonumber \\
&&~\frac{1}{8} \sum_{a=1}^8 [<0| F^{+\mu}_a(x^-,x_T) ~\Phi[x^-,x_T]~a^\dagger_H(P^+,0_T)  a_H(P^+,0_T)~\Phi[0]~  F^+_{\mu a}(0) |0>].
\label{gvf}
\eea

We will present derivations of eqs. (\ref{qnf}) and (\ref{gnf}) in this paper.

The paper is organized as follows. In section II we briefly review the definition of
Collins-Soper fragmentation function in vacuum which is widely used at $pp$, $ep$
and $e^+e^-$ colliders. In section III we give a brief description of Schwinger-Keldysh
closed-time path integral formalism in non-equilibrium QCD. We implement closed-time path
integral formalism in non-equilibrium QCD to Collins-Soper fragmentation
function in section IV. Section V contains conclusion.

\section{ Collins-Soper Fragmentation Function in Vacuum }

In this section we will briefly review Collins-Soper fragmentation function in vacuum
which is widely used at $pp$, $ep$ and $e^+e^-$ colliders. Consider a scalar gluon,
for example, with four momentum $k^\mu$ in vacuum fragmenting to a hadron with four
momentum $P^\mu$. For application to collider experiments it is convenient to use
light cone quantization formalism. The scalar gluon field $\phi(x)$ can be written as
\bea
\phi(x^-,x_T) = \frac{1}{(2\pi)^{d-1}} ~\int \frac{dk^+}{\sqrt{2k^+}} d^{d-2}k_T~[e^{-ik \cdot x} a(k) + e^{ik \cdot x} a^\dagger (k)]_{x^+=0}
\label{phi}
\eea
where $a^\dagger$ and $a$ are the creation and annihilation operators respectively.
$d=4-2\epsilon$ where $3-2\epsilon$ is the space dimension. The single particle parton state is given by
\bea
|k^+, k_T> = a^\dagger (k^+, k_T) |0>, ~~~~~~~~~~~~~{\rm with} ~~~~~~~~~~~~~~~~ a(k^+, k_T) |0> =0
\label{kvec}
\eea
with the normalization
\bea
<k^+, k_T|{k'}^+, {k'}_T> = (2\pi)^{d-1} \delta(k^+-{k'}^+) \delta^{d-2}(k_T -{k'}_T).
\label{norm}
\eea

Note that the correct interpretation of the state $|k>$ is created by an appropriate
Fourier transform of the corresponding field operator and should not be associated
with on-shell condition $k^2=0$ of the massless quark or gluon \cite{george1}.

Consider the inclusive production of a hadron $H$ created in the $out-$state $|H,X>$ from
the parton $a$ in the $in-$state $|k>$ with the probability amplitude
\bea
<H,X|k>.
\eea
The probability distribution $h_k(P)$ of the hadron with momentum $P$ from the parton of
momentum $k$ can be found from the above amplitude. Explicitly
\bea
&& h_k(P) <k|k'>=\sum_X~<k|H,X><H,X|k'>=\sum_X~<k|a^\dagger_H(P)|X><X|a_H(P)|k'> \nonumber \\
&& =<k|a^\dagger_H(P)a_H(P)|k'>
\label{hp2}
\eea
where $a^\dagger_H$ is the creation operator of a hadron $H$. In the light-cone quantization formalism we find
(by using eqs. (\ref{kvec}) and (\ref{norm}))
\bea
&& h(z,P_T) <k^+,k_T|{k'}^+,{k'}_T>=2z(2\pi)^{d-1}D_{H/a}(z,P_T) (2\pi)^{d-1} \delta(k^+-{k'}^+) \delta^{d-2}(k_T -{k'}_T) \nonumber \\
&& =<0|a(k^+,k_T)a^\dagger_H(P^+,P_T)a_H(P^+,P_T)a^\dagger({k'}^+,{k'}_T)|0>
\label{jetf}
\eea
where $D_{H/a}(z,P_T)$ is the fragmentation function and $z=\frac{P^+}{k^+}$
is the longitudinal momentum fraction of hadron $H$ with respect to parton $a$.
Using
\bea
&& (2\pi)^{d-1} <0| \phi(x^-,x_T) a^\dagger_H(P^+,P_T)  a_H(P^+,P_T) \phi(0) |0>
 = \frac{1}{(2\pi)^{d-1}} \int \frac{dk^+}{\sqrt{2k^+}} d^{d-2}k_T \int \frac{d{k'}^+}{\sqrt{2{k'}^+}} d^{d-2}{k'}_T \nonumber \\
 &&~[<0| e^{-ik\cdot x} a(k^+,k_T)a^\dagger_H(P^+,P_T)a_H(P^+,P_T)a^\dagger({k'}^+,{k'}_T)|0>]_{x^+=0} \nonumber \\
\label{fr3}
\eea
we find from eq. (\ref{jetf})
\bea
D_{H/a}(z,P_T)
= \frac{k^+}{z} \int dx^- \frac{d^{d-2}x_T}{(2\pi)^{d-1}}  e^{i{k}^+ x^- - i {k}_T \cdot x_T} <0| \phi(x^-,x_T) a^\dagger_H(P^+,P_T)  a_H(P^+,P_T) \phi(0) |0>.~~~~~
\label{fr6ol}
\eea
It is convenient to rewrite the definition in a form analogous to the definition of the distribution of
partons in a hadron. The transverse momentum is of the parton relative to the hadron rather than vice versa.
For this purpose we make a Lorentz transformation to a frame where the hadron's transverse momentum is
zero:
\bea
&& (P^+, P_T) \rightarrow (P^+,0)  \nonumber \\
&& (k^+,0) \rightarrow (k^+, -P_T/z).
\label{mt}
\eea
Hence we find from eq. (\ref{fr6ol})
\bea
D_{H/a}(z,P_T)
= \frac{k^+}{z} \int dx^- \frac{d^{d-2}x_T}{(2\pi)^{d-1}}  e^{i{k}^+ x^- + i {P}_T \cdot x_T/z}
<0| \phi(x^-,x_T) a^\dagger_H(P^+,0_T)  a_H(P^+,0_T) \phi(0) |0>.~~~~~~~
\label{fr6}
\eea

The $p_T$ integrated fragmentation function is given by
\bea
&& d_{H/a} (z)= \int d^{d-2}P_T D_{H/g}(z,P_T) \nonumber \\
&& =\frac{k^+z^{d-3}}{2\pi}~\int dx^-e^{i{P}^+x^-/z}[
<0|\phi(x^-) a^\dagger_H(P^+,0_T)a_H(P^+,0_T) \phi(0) |0>].
\label{frz2}
\eea

\subsection{ Quark Fragmentation Function }

Following similar steps as above but performing calculation for quark we find the quark fragmentation function
\bea
&& D_{H/q}(z,P_T)
= \frac{1}{2z(2\pi)^{d-1}} \int dx^- d^{d-2}x_T  e^{i{k}^+ x^- + i {P}_T \cdot x_T/z} \nonumber \\
&&~\frac{1}{2}{\rm tr_{Dirac}}~\frac{1}{3}{\rm tr_{color}}[\gamma^+<0| \psi(x^-,x_T) a^\dagger_H(P^+,0_T)  a_H(P^+,0_T) {\bar \psi}(0) |0>]
\label{fr7}
\eea
where $\psi(x)$ is the quark field. The $P_T$ integrated quark fragmentation function becomes
\bea
d_{H/q} (z) = \frac{z^{d-3}}{4\pi}~ \int dx^-  e^{i{P}^+ x^-/z} \frac{1}{2}{\rm tr_{Dirac}}~\frac{1}{3}{\rm tr_{color}}[\gamma^+<0| \psi(x^-) a^\dagger_H(P^+,0_T)  a_H(P^+,0_T) {\bar \psi}(0) |0>].~~
\label{frz3}
\eea

\subsection{ Gluon Fragmentation Function }

Following the above steps but for gluons we find the gluon fragmentation function
\bea
&& D_{H/g}(z,P_T)
= -\frac{1}{2zk^+(2\pi)^{d-1}} \int dx^- d^{d-2}x_T  e^{i{k}^+ x^- + i {P}_T \cdot x_T/z} \nonumber \\
&& ~\frac{1}{8} \sum_{a=1}^8 [<0| F^{+\mu}_a(x^-,x_T) a^\dagger_H(P^+,0_T)  a_H(P^+,0_T)  F^+_{\mu a}(0) |0>]
\label{fr8}
\eea
where
\bea
F^{\mu \nu}_a = \partial^\mu A^\nu_a - \partial^\nu A^\mu_a.
\label{fmn}
\eea
The $P_T$ integrated gluon fragmentation function becomes
\bea
d_{H/g} (z) = -\frac{z^{d-3}}{4\pi k^+}~ \int dx^-  e^{i{P}^+ x^-/z} \sum_{a=1}^8
[<0| F^{+\mu}_a(x^-) a^\dagger_H(P^+,0_T)  a_H(P^+,0_T)  F^+_{\mu a}(0) |0>].
\label{frz4}
\eea

\subsection{ Wilson Lines and Fragmentation Functions }

The quark and gluon fragmentation functions as defined above are not gauge invariant.
Gauge invariant parton fragmentation functions are obtained by incorporating Wilson lines
$\Phi[x]_{ab}$ into the definition of the
quark and gluon fragmentation functions \cite{collins,george,tucci,nayak}. We find
\bea
&& D_{H/q}(z,P_T)
= \frac{1}{2z(2\pi)^{d-1}} \int dx^- d^{d-2}x_T  e^{i{k}^+ x^- + i {P}_T \cdot x_T/z} \nonumber \\
&&~\frac{1}{2}{\rm tr_{Dirac}}~\frac{1}{3}{\rm tr_{color}}[\gamma^+<0| \psi(x^-,x_T) ~\Phi[x^-,x_T]~a^\dagger_H(P^+,0_T)  a_H(P^+,0_T) ~\Phi[0]~{\bar \psi}(0) |0>]~~~~
\label{fr7}
\eea
where $\Phi[x^-,x_T]$ is given by eq. (\ref{wilf}) with $T^{ab}$ in the fundamental
representation of SU(3).

Similarly, incorporating Wilson lines, we find the gauge invariant gluon fragmentation
function
\bea
&& D_{H/g}(z,P_T)
= -\frac{1}{2zk^+(2\pi)^{d-1}} \int dx^- d^{d-2}x_T  e^{i{k}^+ x^- + i {P}_T \cdot x_T/z} \nonumber \\
&& ~\frac{1}{8}
\sum_{a=1}^8 [<0| F^{+\mu}_a(x^-,x_T) ~\Phi[x^-,x_T]~a^\dagger_H(P^+,0_T)  a_H(P^+,0_T) ~\Phi[0]~ F^+_{\mu a}(0) |0>]
\label{fr8}
\eea
where $\Phi[x^-,x_T]$ is given by eq. (\ref{wilf}) with $T^{ab}_c=f^{abc}$ in the adjoint
representation of SU(3).

\section{ Non-equilibrium QCD Using Closed-Time Path Formalism }

Unlike $pp$ collisions, the ground state at RHIC and LHC heavy-ion collisions
(due to the presence of a QCD medium at initial time $t=t_{in}$ (say $t_{in}$=0)
is not a vacuum state $|0>$ any more.
We denote $|in>$ as the initial state of the non-equilibrium QCD
medium at $t_{in}$. The non-equilibrium
distribution function $f(\vec{k})$ of a parton (quark or gluon),
corresponding to such initial state is given by
\bea
<a^\dagger ({\vec k})a({\vec k}')>=<in|a^\dagger ({\vec k})a({\vec k}')|in> = f(\vec{k}) (2\pi)^{d-1} \delta^{(d-1)} ({\vec k} -{\vec k}')
\label{dist}
\eea
where we have assumed space translational invariance at initial time.

Finite temperature field theory formulation is a special case of this when
$f({\vec k}) =\frac{1}{e^{\frac{k_0}{T}} \pm 1}$.

Consider scalar gluons first. In the CTP formalism in non-equilibrium there are four Green's functions
\bea
&& G{++}(x,x') = <in|T\phi (x) \phi (x')|in> = <T\phi (x) \phi (x')>\nonumber \\
&& G{--}(x,x') = <in|{\bar T} \phi (x) \phi (x')|in> = <{\bar T} \phi (x) \phi (x')>\nonumber \\
&& G{+-}(x,x') = <in|\phi (x') \phi (x)|in> = <\phi (x') \phi (x) >\nonumber \\
&& G{-+}(x,x') = <in|\phi (x) \phi (x')|in>= <\phi (x) \phi (x') >
\label{green}
\eea
where $+$($-$) sign corresponds to upper(lower) time branch of the Schwinger-Keldysh
closed-time path \cite{schw,keldysh}. $T$ is the time order product and ${\bar T}$ is the
anti-time order product. The field $\phi(x)$ is in Heisenberg representation. Explicitly
\bea
&& T\phi (x) \phi (x') = \theta(t - t') \phi (x) \phi (x') + \theta(t'-t) \phi (x') \phi (x) \nonumber \\
&& {\bar T} \phi (x) \phi (x') = \theta(t' - t) \phi (x) \phi (x') + \theta(t-t') \phi (x') \phi (x).
\label{ttbar}
\eea

At initial time $t=t_{in}=0$ the Heisenberg picture coincide with the Schrodinger and interaction
pictures. We write
\bea
\phi(x) =  \int \frac{d^{d-1} k}{(2\pi )^{d-1} \sqrt{ 2k^0}}
[a({\vec k})e^{-ik \cdot x} + a^\dagger ({\vec k}) e^{i k \cdot x}]
\label{phi}
\eea
where $a^\dagger ({\vec k})$ and $a ({\vec k})$ are creation and annihilation operators
respectively.
The commutation relations are given by
\bea
&& [a({\vec k}), a^\dagger ({\vec k}') ]_{t=0} = (2\pi)^{d-1} \delta^{(d-1)} ({\vec k} -{\vec k}'), \nonumber \\
&& [a({\vec k}), a ({\vec k}') ]_{t=0} =[a^\dagger({\vec k}), a^\dagger ({\vec k}') ]_{t=0} =0.
\label{comm}
\eea
We assume space-translational invariance at initial time and find
\bea
[G_{ij} (x, x')]_{t=0} = [\int d^dk G_{ij} (k) e^{-i k \cdot (x - x')}]_{t=0}
\label{fourier}
\eea
where $i,j$ are $+,-$. Explicitly
\bea
&& G_{-+}(x,x') = <\phi (x) \phi (x') > = \int \frac{d^{d-1}k}{\sqrt{2k^0}(2\pi )^{d-1}}
\int \frac{d^{d-1}k'}{\sqrt{2{k'}^0}(2\pi )^{d-1}} \nonumber \\
&&~<in|[a({\vec k})e^{-ik \cdot x} + a^\dagger ({\vec k}) e^{i k \cdot x}]
[a({\vec k}')e^{-ik' \cdot x'} + a^\dagger ({\vec k}') e^{i k' \cdot x'}] |in>.
\label{gmp}
\eea
Using Bogolyubov transformation we can set
\bea
<in|a({\vec k})a({\vec k}')|in> = <in|a^\dagger ({\vec k})a^\dagger ({\vec k}')|in> =0.
\label{bob}
\eea
By using eqs. (\ref{dist}), (\ref{comm}) and (\ref{bob}) in eq. (\ref{gmp}) we find
\bea
&& [G_{-+}(x,x')]_{t=0} = [\int \frac{d^{d-1}k}{{2k^0}(2\pi )^{d-1}}
[[1 + f({\vec k})]e^{-ik \cdot (x - x')} + f({\vec k}) e^{i k \cdot (x - x')}]_{t=0} \nonumber \\
&& = [\int \frac{d^dk}{(2\pi )^{d-1}} \delta(k^2)e^{-ik \cdot (x - x')}
[\theta(k_0) [1 + f({\vec k})] + \theta(-k_0) f(-{\vec k})]_{t=0} = [\int d^dk G_{-+}(k) e^{-ik \cdot (x - x')}]_{t=0}. \nonumber \\
\label{gmp1}
\eea
Similarly
\bea
&&~[G_{+-}(k)]_{t=0}=\delta(k^2)[\theta(-k_0)+\theta(k_0) f({\vec k})+\theta(-k_0) f(-{\vec k})] \nonumber \\
&&~[G_{++}(k)]_{t=0} = \frac{1}{k^2 +i\epsilon} +\delta(k^2)[\theta(k_0) f({\vec k})+\theta(-k_0) f(-{\vec k})] \nonumber \\
&&~[G_{--}(k)]_{t=0} = \frac{-1}{k^2 -i\epsilon} + \delta(k^2)[\theta(k_0) f({\vec k})+\theta(-k_0) f(-{\vec k})].
\label{gt0}
\eea

\subsection{ Quarks in non-Equilibrium }

The non-equilibrium (massless) quark propagator at initial time $t=t_{in}$ is given by
(suppression of color indices are understod)
\bea
G(k)_{ij}=\displaystyle{\not}k \left ( \begin{array}{cc}
\frac{1}{k^2+i\epsilon}+2\pi \delta(k^2) f_q(\vec{k}) & -2\pi \delta(k^2)\theta(-k_0)+2\pi \delta(k^2) f_q(\vec{k}) \\
-2\pi \delta(k^2)\theta(k_0)+2\pi \delta(k^2) f_q(\vec{k}) & -\frac{1}{k^2-i\epsilon}+2\pi \delta(k^2) f_q(\vec{k})
\end{array} \right )
\eea
where where $i,j= +,-$ and $f_q(\vec{k})$ is the arbitrary non-equilibrium distribution
function of quark.

\subsection{ Gluons in Non-Equilibrium }

We work in the frozen ghost formalism \cite{greiner,cooper} where the non-equilibrium
gluon propagator at initial time $t=t_{in}$ is given by (the suppression of color
indices are understood)
\bea
G^{\mu \nu}(k)_{ij} = -i[g^{\mu \nu} +  (\alpha -1) \frac{k^\mu k^\nu}{k^2}] ~G^{\rm vac}_{ij}(k) -iT^{\mu \nu}G^{\rm med}_{ij}(k)
\label{gpm}
\eea
where $i,j= +,-$. The transverse tensor is given by
\bea
T^{\mu \nu} (k)=g^{\mu \nu} -\frac{(k \cdot u)(u^\mu k^\nu + u^\nu k^\mu)-k^\mu k^\nu -k^2u^\mu u^\nu}{(k \cdot u)^2 -k^2}
\label{tmn}
\eea
with the flow velocity of the medium $u^\mu$. $G^{\mu \nu}_{ij}(k)$ are the usual vacuum propagators of the gluon
\bea
G^{\rm vac}_{ij}(k)=
\left ( \begin{array}{cc}
\frac{1}{k^2+i\epsilon} & -2\pi \delta(k^2)\theta(-k_0) \\
-2\pi \delta(k^2)\theta(k_0) & -\frac{1}{k^2-i\epsilon}
\end{array} \right )
\eea
and the medium part of the propagators are given by
\bea
G^{\rm med}_{ij}(k)= 2\pi \delta(k^2) f_g(\vec{k})
\left ( \begin{array}{cc}
1 & 1 \\
1 & 1
\end{array} \right ).
\eea

\section{ Fragmentation Function in Non-Equilibrium QCD Using Closed-Time Path Formalism }

For simplicity, let us consider the scalar gluon first in the light cone quantization formalism.
Generalizing the vacuum analysis in \cite{collins} we define the state $|k^+,k_T>$ of the fragmenting
gluon in non-equilibrium QCD medium at initial time $x^+=x^+_{in}$ (say at $x^+_{in}=0$)
\bea
|k^+, k_T> = a^\dagger (k^+, k_T) |in>.
\label{kt}
\eea
The non-equilibrium distribution function $f(k^+,k_T)$ of the fragmenting gluon is given by
\bea
<in|a^\dagger (k^+,k_T)a({k^\prime}^+,k^\prime_T)|in> = f(k^+,k_T) (2\pi)^{d-1} \delta(k^+-{k^\prime}^+) \delta^{(d-2)} ( k_T - k^\prime_T)
\label{distl}
\eea
where we have assumed the space ($x^-$ and $x_T$) translational invariance at initial time $x^+=x^+_{in}$.
The commutation relations are given by
\bea
&& [a(k^+,k_T), a^\dagger ({k^\prime}^+, k^\prime_T) ]_{x^+=0} = (2\pi)^{d-1} \delta(k^+-{k^\prime}^+) \delta^{(d-2)} (k_T - k^\prime_T), \nonumber \\
&& [a(k^+,k_T), a ({k^\prime}^+,k^\prime_T) ]_{x^+=0} =[a^\dagger( k^+,k_T), a^\dagger ({ k^\prime }^+, k'_T) ]_{x^+=0} =0
\label{comml}
\eea
By using eqs. (\ref{dist}) and (\ref{comm}) we find
\bea
&& <k^+, k_T|{k'}^+, {k'}_T> = <in|a(k^+, k_T)a^\dagger({k'}^+, {k'}_T)|in> \nonumber \\
&& =(2\pi)^{d-1} \delta(k^+-{k'}^+) \delta^{d-2}(k_T -{k'}_T) [1+f(k^+,k_T)].
\label{norml}
\eea

Consider the inclusive production of hadron $H$ created in the $out-$state $|H,X>$ from
a scalar gluon in non-equilibrium in the initial state $|k>$ with the probability amplitude
\bea
<H,X|k>.
\eea
$X$ being other outgoing final state particles.
Similar to the vacuum case of Collins-Soper fragmentation function,
the correct interpretation of the above state $|k>$ is created by an appropriate
Fourier transform of the corresponding field operator and should not be associated
with on-shell condition $k^2=0$ of the massless quark or gluon \cite{george1}.
The distribution $h_k(P)$ of the hadron $H$ with momentum $P$ from the parton of
momentum $k$ can be found from the above amplitude. We find
\bea
\sum_X~<k,k_T|H,X><H,X|{k^+}',k'_T>=h_k(P) <k^+,k_T|{k^+}',k'_T>.
\label{hp1l}
\eea
For the left hand side we write
\bea
&& \sum_X~<k^+,k_T|H,X><H,X|{k^+}',k'_T>= \sum_X~<k^+,k_T|a^\dagger_H(P)|X><X|a_H(P)|{k^+}',k'_T> \nonumber \\
&& =<k^+,k_T|a^\dagger_H(P)a_H(P)|{k^+}',k'_T>.
\label{hp2l}
\eea
Equating eqs. (\ref{hp1l}) and (\ref{hp2l}) and by using eq. (\ref{kt}) we find
\bea
&& <in|a(k^+,k_T)a^\dagger_H(P^+,P_T)a_H(P^+,P_T)a^\dagger({k'}^+,{k'}_T)|in> \nonumber \\
&& =2z(2\pi)^{d-1}D_{H/a}(z,P_T)<in|a(k^+,k_T)a^\dagger({k'}^+,{k'}_T)|in>.
\label{fr1la}
\eea
This expression is exactly similar to that of the jet fragmentation function
in vacuum as given in eq. (\ref{jetf}) except that the vacuum expectation is
replaced by medium average at the initial time $x^+=x^+_{in}$. Using eqs. (\ref{kt})
and (\ref{distl}) we find
\bea
<in|a(k^+,k_T)a^\dagger_H(P^+,P_T)a_H(P^+,P_T)a^\dagger({k'}^+,{k'}_T)|in>
=2z(2\pi)^{d-1}D_{H/a}(z,P_T)~[1+f(k^+,k_T)]. \nonumber \\
\label{fr1lal}
\eea
From eq. (\ref{phi}) we obtain
\bea
&& (2\pi)^{d-1} <in| \phi(x^-,x_T) a^\dagger_H(P^+,P_T)  a_H(P^+,P_T) \phi(0) |in>
 = \frac{1}{(2\pi)^{d-1}} \int \frac{dk^+}{\sqrt{2k^+}} d^{d-2}k_T \int \frac{d{k'}^+}{\sqrt{2{k'}^+}} d^{d-2}{k'}_T \nonumber \\
  &&~[<in| e^{-ik\cdot x} a(k^+,k_T)a^\dagger_H(P^+,P_T)a_H(P^+,P_T)a^\dagger({k'}^+,{k'}_T)|in>]_{x^+=0}. \nonumber \\
\label{fr3m}
\eea

Using this in eq. (\ref{fr1lal}) we find the expression of the fragmentation function in non-equilibrium QCD
\bea
&& D_{H/a}(z,P_T)= \frac{k^+}{z~[1+f(k^+,k_T)]} \int dx^- \frac{d^{d-2}x_T}{(2\pi)^{d-1}}  e^{i{k}^+ x^- - i {k}_T \cdot x_T} \nonumber \\
&& <in| \phi_a(x^-,x_T) a^\dagger_H(P^+,P_T)  a_H(P^+,P_T) \phi_a(0) |in>. \nonumber \\
\label{fr6ollp}
\eea
From eq. (\ref{mt}) we find
\bea
&& D_{H/a}(z,P_T)= \frac{k^+}{z~[1+f(k^+,k_T)]} \int dx^- \frac{d^{d-2}x_T}{(2\pi)^{d-1}}  e^{i{k}^+ x^- + i {P}_T \cdot x_T/z} \nonumber \\
&& <in| \phi_a(x^-,x_T) a^\dagger_H(P^+,0_T)  a_H(P^+,0_T) \phi_a(0) |in>. \nonumber \\
\label{fr6oll}
\eea
In the above expression, $f(k^+,k_T)$ is the non-equilibrium distribution function of the fragmenting gluon
at initial time $x^+=x^+_{in}$ and $|in>$ is the initial state of the non-equilibrium QCD medium in the Schwinger-Keldysh $in-in$ closed-time path formalism.

\subsection{ Quark Fragmentation Function in Non-Equilibrium }

Following the above steps, but for quarks, we find the quark fragmentation function
in non-equilibrium QCD
\bea
&& D_{H/q}(z,P_T)
= \frac{1}{2z~[1+f_q(k^+,k_T)]} \int dx^- \frac{d^{d-2}x_T}{(2\pi)^{d-1}}  e^{i{k}^+ x^- + i {P}_T \cdot x_T/z} \nonumber \\
 &&~\frac{1}{2}{\rm tr_{Dirac}}~\frac{1}{3}{\rm tr_{color}}[\gamma^+<in| \psi(x^-,x_T) a^\dagger_H(P^+,0_T)  a_H(P^+,0_T) {\bar \psi}(0)  |in>
\label{qnon}
\eea
where $f_q(k^+,k_T)$ is the non-equilibrium distribution function of the fragmenting quark at initial time.

\subsection{ Gluon Fragmentation Function in Non-Equilibrium }

For gluons we consider frozen ghost formalism described above (\ref{gpm}).
Hence all the analysis of the above can be applied.
Carrying out the similar algebra as above we find the gluon fragmentation function
\bea
&& D_{H/g}(z,P_T)
= -\frac{1}{2zk^+~[1+f_g(k^+,k_T)]} \int dx^- \frac{d^{d-2}x_T}{(2\pi)^{d-1}}  e^{i{k}^+ x^- + i {P}_T \cdot x_T/z} \nonumber \\
&&~\frac{1}{8} \sum_{a=1}^8 [<in| F^{+\mu}_a(x^-,x_T) a^\dagger_H(P^+,0_T)  a_H(P^+,0_T)  F^+_{\mu a}(0) |in>]
\label{gnon}
\eea
where $f_g(k^+,k_T)$ is the non-equilibrium distribution function of the fragmenting gluon at initial time.

\subsection{ Wilson Lines }

The above parton to hadron fragmentation function definition in non-equilibrium QCD is not gauge invariant.
To make it gauge invariant we need to incorporate Wilson lines.
Incorporating the Wilson lines (eq. (\ref{wilf})) into the definition of the fragmentation function
eq. (\ref{qnon}) we find the gauge invariant quark fragmentation function in non-equilibrium QCD
\bea
&& D_{H/q}(z,P_T)
= \frac{1}{2z~[1+f_q(k^+,k_T)]} \int dx^- \frac{d^{d-2}x_T}{(2\pi)^{d-1}}  e^{i{k}^+ x^- + i {P}_T \cdot x_T/z} \nonumber \\
 &&~\frac{1}{2}{\rm tr_{Dirac}}~\frac{1}{3}{\rm tr_{color}}[\gamma^+<in| \psi(x^-,x_T) ~\Phi[x^-,x_T]~a^\dagger_H(P^+,0_T)  a_H(P^+,0_T) ~\Phi[0]~{\bar \psi}(0)  |in>] \nonumber \\
\label{qnonf}
\eea
where $f_q(k^+,k_T)$ is the non-equilibrium distribution function of the fragmenting
quark at initial time $x^+=x^+_{in}$. $|in>$ is the initial state of the non-equilibrium QCD
medium in the Schwinger-Keldysh $in-in$ closed-time path formalism. This reproduces eq. (\ref{qnf}).

Incorporating the Wilson lines (eq. (\ref{wilf})) into the definition of the fragmentation function
eq. (\ref{gnon}) we find the gauge invariant gluon fragmentation function in non-equilibrium QCD
\bea
&& D_{H/g}(z,P_T)
=- \frac{1}{2zk^+~[1+f_g(k^+,k_T)]} \int dx^- \frac{d^{d-2}x_T}{(2\pi)^{d-1}}  e^{i{k}^+ x^- + i {P}_T \cdot x_T/z} \nonumber \\
&&~\frac{1}{8} \sum_{a=1}^8 [<in| F^{+\mu}_a(x^-,x_T) ~\Phi[x^-,x_T]~a^\dagger_H(P^+,0_T)  a_H(P^+,0_T)~\Phi[0]~  F^+_{\mu a}(0) |in>]
\label{gnonf}
\eea
where $f_g(k^+,k_T)$ is the non-equilibrium distribution function of the fragmenting
gluon at initial time $x^+=x^+_{in}$. $|in>$ is the initial state of the non-equilibrium QCD
medium in the Schwinger-Keldysh $in-in$ closed-time path formalism. This reproduces eq. (\ref{gnf}).

It may be possible to include final state interactions
with the medium via modification of the Wilson lines in this definition of the non-equilibrium
fragmentation function, similar to the $p_T$ distribution of the
parton distribution function studied in \cite{beli}.

\section{Conclusions}

In this paper we have implemented closed-time path integral formalism in non-equilibrium
QCD to the definition of Collins-Soper fragmentation function. We have
considered a high $p_T$ parton in QCD medium at initial time $\tau_0$ with arbitrary non-equilibrium (non-isotropic) distribution function $f(\vec{p})$ fragmenting to hadron.
We have formulated parton to hadron fragmentation function in non-equilibrium QCD in the
light-cone quantization formalism. This may be relevant to study hadron production from quark-gluon plasma at RHIC
and LHC. It may be possible to include final state interactions
with the medium via modification of the Wilson lines in this definition of the non-equilibrium
fragmentation function, similar to the $p_T$ distribution of the
parton distribution function studied in \cite{beli}. This will be the subject of a future
analysis.

\acknowledgements

I thank Geoff Bodwin, Fred Cooper, Jianwei Qiu and George Sterman for useful
discussions. This work is supported in part by the U.S. Department of Energy
under Grant no. DE-FG02-01ER41195.

\end{document}